\documentclass[%
 reprint,
superscriptaddress,
showpacs,preprintnumbers,
 amsmath,amssymb,
 aps,nolinenumbers
]{revtex4-2}
\usepackage{ulem}
\usepackage{ragged2e}
\usepackage{caption}
\usepackage{graphicx}
\usepackage{dcolumn}
\usepackage{bm}
\usepackage{xcolor}
\usepackage{braket}
\usepackage{lineno}
\usepackage{url}
\newcommand{\etal}{{\it et al.}}
\newcommand{\ie}{{\it i.e.\ }}
\newcommand{\CsVSb}{{CsV$_{3}$Sb$_{5}$}}

\begin{document}
\captionsetup{justification=centerlast , labelsep = space,labelfont=bf,format=plain,singlelinecheck=off}

\title{Nodeless superconductivity in kagome metal CsV$_{3}$Sb$_{5}$ \\with and without time reversal symmetry breaking}

\author{Wei~Zhang}
\author{Xinyou~Liu}
\author{Lingfei~Wang}
\author{Chun~Wai~Tsang}
\author{Zheyu~Wang}
\author{Siu~Tung~Lam}
\author{Wenyan~Wang}
\author{Jianyu~Xie}
\affiliation{Department of Physics, The Chinese University of Hong Kong, Shatin, Hong Kong, China}
\author{Xuefeng~Zhou}
\author{Yusheng~Zhao}
\author{Shanmin~Wang}
\affiliation{Department of Physics, Southern University of Science and Technology, Shenzhen, Guangdong, China}
\author{Jeff~Tallon}
\affiliation{Robinson Institute, Victoria University of Wellington, P.O. Box 600, Wellington, New Zealand}
\author{Kwing~To~Lai$^{*}$}
\affiliation{Department of Physics, The Chinese University of Hong Kong, Shatin, Hong Kong, China} 
\affiliation{Shenzhen Research Institute, The Chinese University of Hong Kong, Shatin, Hong Kong, China}
\author{Swee~K.~Goh$^{\dagger}$}
\affiliation{Department of Physics, The Chinese University of Hong Kong, Shatin, Hong Kong, China}

\date{\today}

\begin{abstract}

\noindent The kagome metal \CsVSb\ features an unusual competition between the charge-density-wave (CDW) order and superconductivity. Evidence for time-reversal symmetry breaking (TRSB) inside the CDW phase has been accumulating. Hence, the superconductivity in \CsVSb\ emerges from a TRSB normal state, potentially resulting in an exotic superconducting state. To reveal the pairing symmetry, we first investigate the effect of nonmagnetic impurity. Our results show that the superconducting critical temperature is insensitive to disorder, pointing to conventional $s$-wave superconductivity. Moreover, our measurements of the self-field critical current ($I_{c,sf}$), which is related to the London penetration depth, also confirm conventional $s$-wave superconductivity with strong coupling. Finally, we measure $I_{c,sf}$ where the CDW order is removed by pressure and superconductivity emerges from the pristine normal state. Our results show that $s$-wave gap symmetry is retained, providing strong evidence for the presence of conventional $s$-wave superconductivity in \CsVSb\
irrespective of the presence of the TRSB.\\

\end{abstract}

\maketitle

Kagome metals AV$_3$Sb$_5$ (A=K, Rb, Cs) have been heavily studied recently due to their exotic properties including non-trivial topology, anomalous Hall effect (AHE), and interesting interplay between superconductivity and unconventional charge density wave (CDW)~\cite{Ortiz2019,Ortiz2020,Du2021,Chen2021a,Li2021,Liang2021,Zhao2021,Liu2021,Hu2021,Uykur2021,Zhou2021,Jiang2021,Kang2022,Kang2022a,Wu2022,Lou2022,Yu2021b,Yang2020,Yu2021,Yu2022}. Among the three compounds, \CsVSb~possesses the highest $T_c \sim$ 2.7 K with a second-order CDW phase transition occurring at $T_{\rm CDW} \sim$ 90~K \cite{Ortiz2019,Ortiz2020}. From the zero-field muon spin relaxation (ZF-$\mu$SR) and magneto-optic polar Kerr effect measurements, evidence of time-reversal symmetry breaking (TRSB) has been detected in the CDW phase. Concurrently, anomalous Hall effect without local moments occurs~\cite{Yu2021,Yu2021b,Yu2021c,Hu2022}. To explain the observed AHE and TRSB, a chiral flux phase (CFP) of CDW has been proposed~\cite{Feng2021}. Hence, the superconducting state in \CsVSb~emerges from a TRSB normal state, potentially resulting in an exotic superconducting ground state.

The pairing symmetry can shed light on the unconventional nature of the superconductivity.
However, the pairing symmetry of \CsVSb\ remains controversial based on existing experimental results. Scanning tunneling microscopy (STM) has detected a V-shape density of states, indicating a gapless superconductivity \cite{Liang2021,Xu2021,Chen2021b}. Furthermore, thermal conductivity shows a finite residual linear term in the 0~K limit, lending support to a nodal superconducting gap~\cite{Zhao2021a}. On the other hand, the magnetic penetration depth revealed by both tunnel diode oscillator (TDO) and muon spin rotation ($\mu$SR) experiments suggest nodeless superconductivity in \CsVSb~\cite{Gupta2022,Gupta2022a,Duan2021,Shan2022,Roppongi2022}.
Moreover, from the spin-lattice relaxation measurement, a finite Hebel-Slichter coherence peak is observed just below $T_c$, indicating a conventional $s$-wave pairing~\cite{Mu2021}.

Whether or not TRSB has an influence on the pairing symmetry needs to be clarified urgently. One approach is to remove the CDW state completely, and investigate the pristine superconducting phase. This total suppression of the CDW phase can be achieved by applying a hydrostatic pressure greater than $\sim$20~kbar~\cite{Yu2021,Chen2021a}. Thus, a careful examination of the superconducting ground state of \CsVSb\ without the complication due to TRSB can be performed, and this forms the major theme of this manuscript. To accomplish this objective, a probe that can detect the superconducting gap under pressure is needed.

Recently, the superfluid density has been shown to be related to the self-field critical current density ($J_{c,sf}$), \ie the transport critical current density in the absence of an external magnetic field~\cite{Talantsev2015, Talantsev2017, Liu2022}. Thus, by measuring the temperature dependence of the critical current, the gap symmetry and the coupling strength can be extracted. Measurement of $J_{c,sf}$ has been demonstrated under pressure~\cite{Seo2015,Semenok2022}. Therefore, the examination of $J_{c,sf}$ provides a novel route to probe the superconducting gap and superfluid density at any pressure.

Apart from the self-field critical current, the effect of nonmagnetic impurities can also give information on the superconducting gap symmetry. In $s$-wave superconductors without sign change of the gap, Anderson’s theorem dictates that Cooper pairs are not destroyed by nonmagnetic impurities and hence, $T_c$ is more robust against the disorder level~\cite{Anderson1959,Hirschfeld2016}. However, if the gap is formed by portions with different signs or there are nodes in the gap, such as an $s_{\pm}$ state or a $d$-wave state, nonmagnetic impurities will be pair-breaking and suppress $T_c$ rapidly~\cite{Anderson1959,Hirschfeld2016,Mackenzie1998,Chen2016,Mizukami2014,Prozorov2014}.

In this article, we explore the nature of the superconducting gap in \CsVSb\ with and without TRSB. At ambient pressure, where TRSB is present, we measure the $T_c$ of a large number of crystals with varying residual resistivity ratios (RRRs). Scanning tunneling microscopy detected the presence of Cs/Sb vacancies or V defect~\cite{Xu2021,Nie2022}. Therefore, the different RRR values could originate from different concentrations of vacancies and defect. Crucially, these are nonmagnetic impurities, providing the avenue to investigate their effect on $T_c$. Next, we probe the superconducting state by $J_{c,sf}$. The insensitivity of $T_c$ to disorder and the $T$-dependence of $J_{c,sf}$ both indicate a conventional $s$-wave superconductivity. To investigate the role of the TRSB CDW phase on the superconductivity, we further detect the critical current under high pressures where the CDW phase is totally suppressed. The superconductivity which emerges from the pristine phase also follows the nodeless $s$-wave gap symmetry. Our results show that TRSB in the CDW phase does not modify the nodeless property of the superconducting gap.

\begin{figure}[!t]\centering
      \resizebox{9cm}{!}{
         \includegraphics{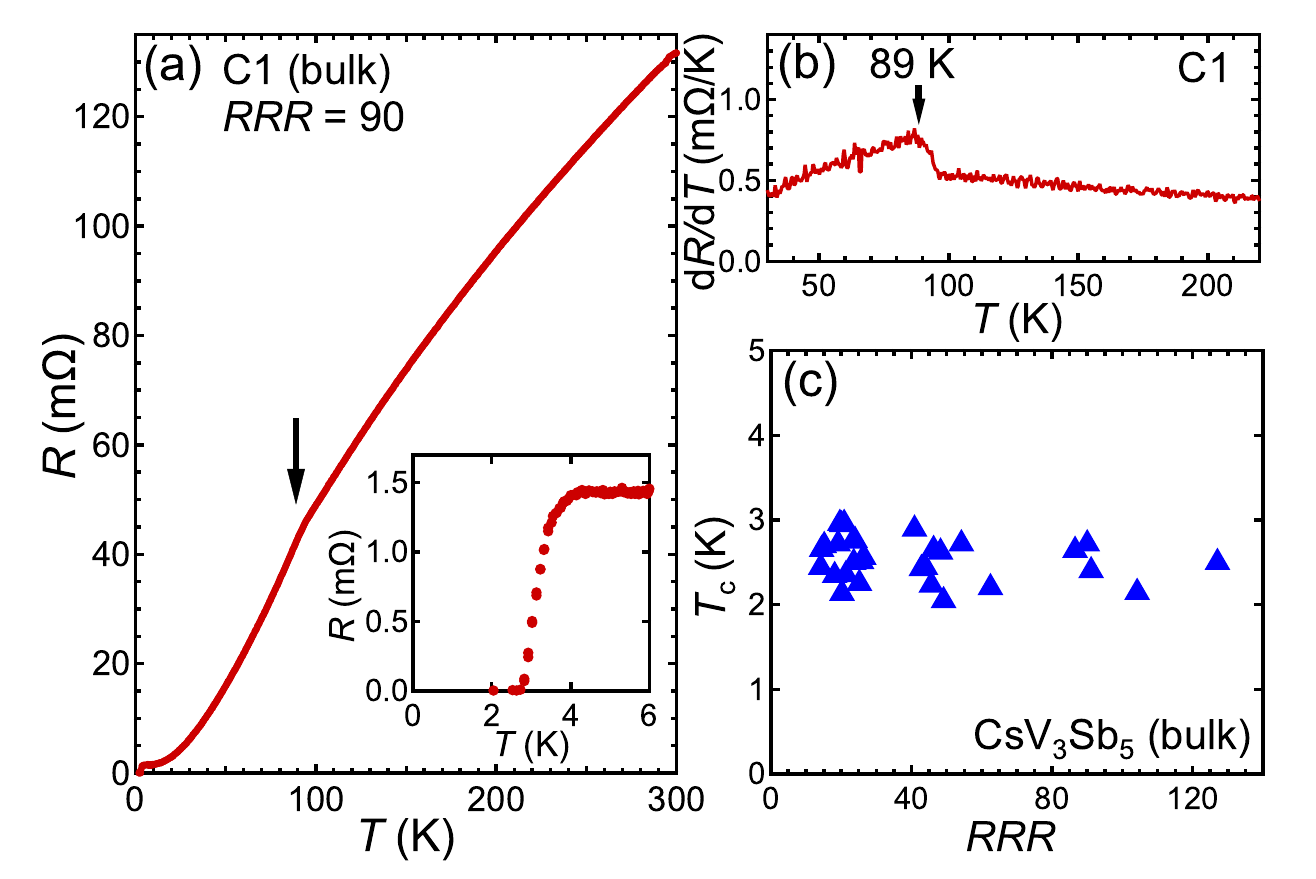}}                				
              \caption{\label{fig1}  
               \textbf{(a)} Temperature dependence of electrical resistance of the bulk \CsVSb\ (sample C1) with a RRR of 90. The inset shows the superconducting transition. \textbf{(b)} Temperature dependence of ${\rm d}R/{\rm d}T$, displaying a sharp peak at $T_{\rm CDW}$. \textbf{(c)} $T_c$ of 30 bulk samples \leftline{with different RRR values.}
              }
\end{figure}

\begin{figure}[!t]\centering
      \resizebox{9cm}{!}{
 \includegraphics{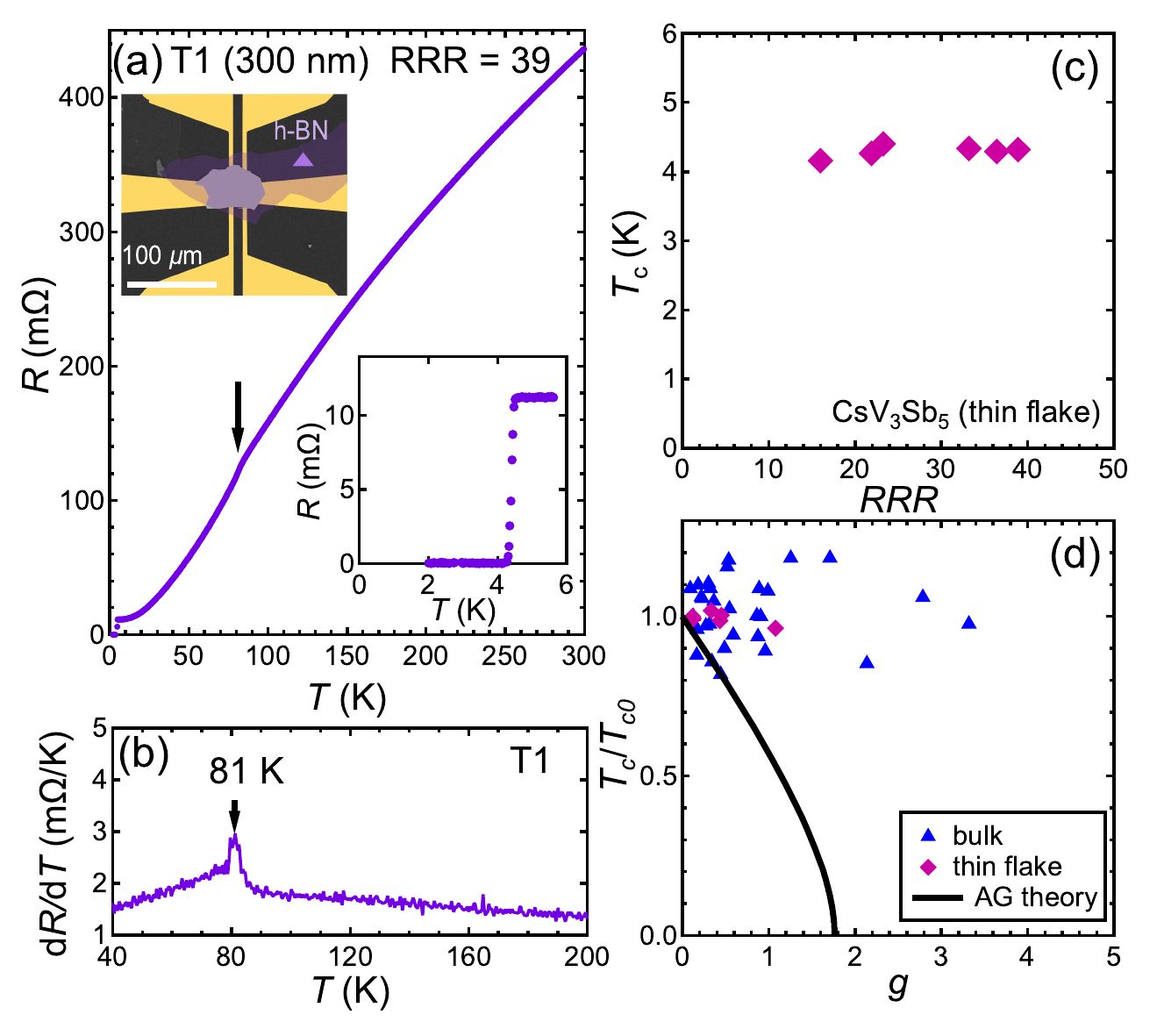}}       \caption{\label{fig2} 
 \textbf{(a)} Temperature dependence of electrical resistance of a thin flake of \CsVSb\ (T1). The upper inset is the scanning electron microscope (SEM) image of a \CsVSb\ thin flake sitting on diamond anvil pre-patterned with electrodes. The sample is covered with h-BN. False colors are used for illustration. The lower inset shows the superconducting transition. \textbf{(b)} Temperature dependence of ${\rm d}R/{\rm d}T$, displaying a sharp peak at $T_{\rm CDW}$. \textbf{(c)} $T_c$ of thin flakes with various RRR values. \textbf{(d)} $T_c$ of \CsVSb~as a function of the dimensionless scattering rate $g$. The solid line represents the suppression of $T_c$ 
\leftline{expected in the Abrikosov-Gor’kov (AG) theory.}
}            
\end{figure}

Figure~\ref{fig1}(a) shows the temperature dependence of the electrical resistance $R(T)$ for one of the single-crystalline \CsVSb~samples in bulk form. On cooling, $R(T)$ decreases and an anomaly appears at around 89~K. Correspondingly, a peak appears in ${\rm d}R/{\rm d}T$, as displayed in Fig.~\ref{fig1}(b), which is consistent with the reported CDW transition. With further cooling, $R(T)$ shows the superconducting transition with $T_c\sim$2.8~K (see the lower inset in Fig.~\ref{fig1}(a)). We have taken the temperature at which the resistance reaches zero as $T_c$. The RRR (defined as $R(300~{\rm K})$/$R(5~{\rm K})$) is 90 for sample C1. Next, we study 30 \CsVSb~samples in bulk form (see Supporting Information S1 for additional $\rho-T$ curves). As shown in Fig.~\ref{fig1}(c), although the RRR spans a large range from 14 to 127, $T_c$ in \CsVSb~is clearly independent of the RRR values. Therefore, $T_c$ is insensitive to disorder, suggesting the absence of nodes in the superconducting gap~\cite{Mackenzie1998,Chen2016}. We note that the RRR dependence of $T_{\rm CDW}$ is also weak (see Supporting Information S1).

\begin{figure*}[!t]\centering
      \resizebox{18.5cm}{!}{
           \includegraphics{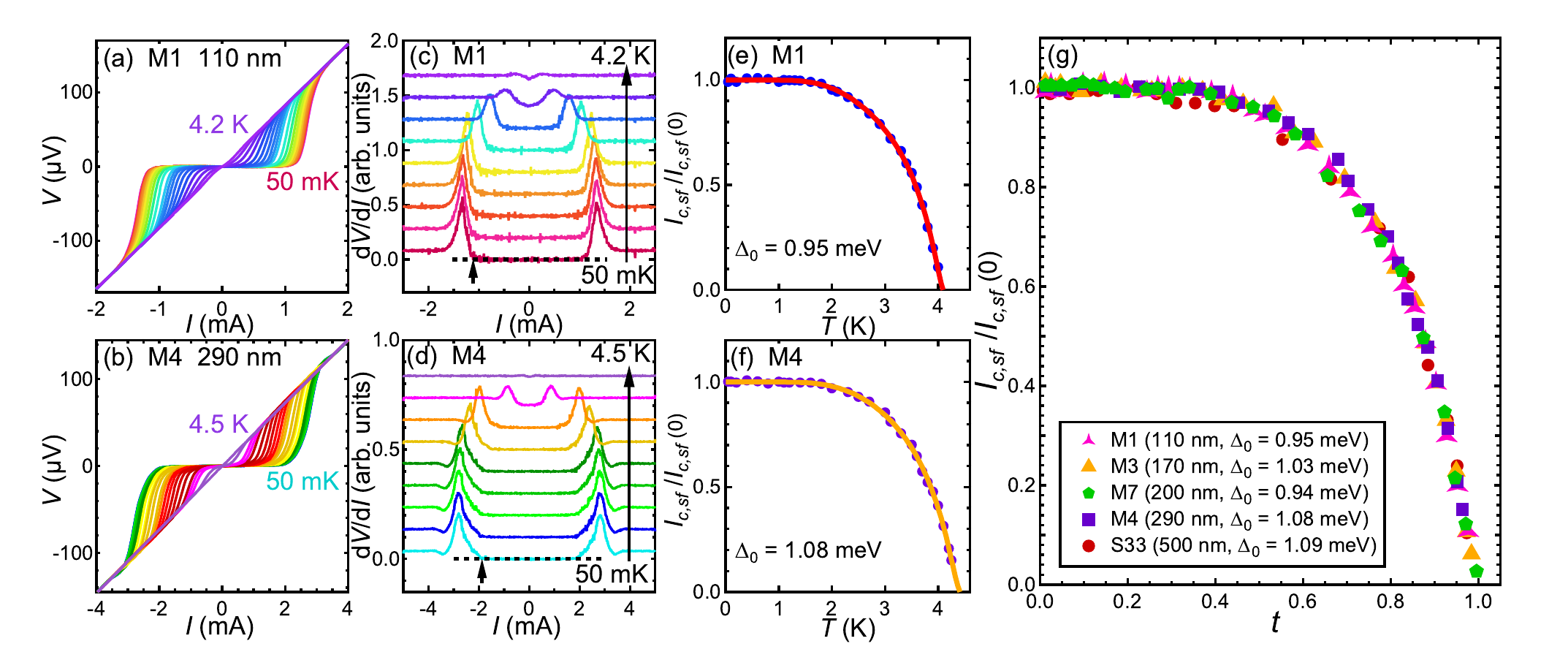}}
 
           \caption{\label{fig3} 
            $V$-$I$ characteristics for thin flake \textbf{(a)} M1 (110 nm) and \textbf{(b)} M4 (290 nm). The calculated first derivative of $V(I)$, d$V$/d$I$, of \textbf{(c)} M1 and \textbf{(d)} M4. The short black arrows indicate the onset of deviation from d$V$/d$I=0$, which we define as the critical current $I_c$. The long arrows show the direction of increasing temperature. Temperature dependence of critical current normalized by critical current at 0~K limit $I_{c,sf}(0)$ of \textbf{(e)} M1 and \textbf{(f)} M4. The solid curves are the single $s$-wave gap fits. \textbf{(g)} $I_{c,sf}(T)/I_{c,sf}(0)$ as a function of the reduced temperature 
           \leftline{$t=T/T_c$ for all measured thin flakes.}
}            
\end{figure*}

We have also studied \CsVSb\ in the form of thin flakes. As shown in the upper inset of Fig.~\ref{fig2}(a), we conduct electrical transport measurements on the flake placed on a diamond substrate pre-patterned with electrodes. The diamond substrate provides an ideal platform to ensure that the flake is thermally anchored to the cold head. We have adopted a similar configuration to measure the Shubnikov-de Haas effect of thin \CsVSb~\cite{Zhang2022}. As we reported in Ref.~\cite{Zhang2022}, $T_c$ is enhanced and the superconducting transition becomes sharper in the thin flake. The higher $T_c$ may result from a possible orbital selective hole doping mechanism~\cite{Zhang2022}. Here, we study six \CsVSb\ thin flakes to investigate the RRR dependence (see Supporting Information S1 for more $\rho-T$ curves). As shown in Fig.~\ref{fig2}(c), $T_c$ is still independent of RRR and it is less scattered about the average value compared with the bulk counterpart. Note that to rule out possible influence of the thickness dependence, all the thin flake samples used in Fig.~\ref{fig2}(c) are around 250~nm.

To quantify the rate at which $T_c$ can be suppressed by impurities, we introduce a dimensionless scattering rate, defined as~\cite{De2018}
\begin{equation}
g=\frac{\hbar}{k_B\mu_0}\frac{\rho_0}{T_{c0}\lambda^2}
\end{equation}
where $\mu_0$ is the vacuum permeability, $k_B$ is Boltzmann constant, $\rho_0$ is the residual resistivity and $\lambda$ is the London penetration depth. In this study, we use the resistivity values at 5~K for $\rho_0$ for each sample and we take $\lambda=$ 450~nm~\cite{Gupta2022}. $T_{c0}$ is the transition temperature in the clean limit. Hence, we take the $T_c$ of the largest RRR sample as $T_{c0}$: $T_{c0}=2.5$~K for the bulk sample and $T_{c0}=4.3$~K for the thin flake. Figure~\ref{fig2}(d) shows $T_c/T_{c0}$ against $g$ for all samples (symbols). Also shown on the figure is a theoretical curve based on Abrikosov-Gor’kov theory, which describes the suppression of $T_c$ for a superconductor with a sign-changing gap when nonmagnetic impurities are introduced, or for a conventional $s$-wave superconductor in the presence of magnetic impurities. As can be seen in Fig.~\ref{fig2}(d), $T_c$ in \CsVSb\ is not sensitive to disorder even though $g$ has spanned a large range, unambiguously pointing to a conventional $s$-wave  superconductivity scenario.

To corroborate the proposal of the nodeless superconductivity in \CsVSb, we further conduct critical current measurements on the thin flakes of \CsVSb. Figures~\ref{fig3}(a) and \ref{fig3}(b) show the voltage-current ($V$-$I$) curves at various temperatures in two representative thin flake samples M1 (thickness, $2b=170$~nm) and M4 ($2b=290$~nm). At a fixed temperature, a pulsed current is applied perpendicular to the cross-section of the thin flakes. For a typical trace, a drastic increase of the voltage is recorded when the current exceeds a threshold, indicating a recovery from the superconducting state to the normal state. We take the current when d$V$/d$I$ first deviates from zero to be the critical current, see the short arrows in Figs.~\ref{fig3}(c) and \ref{fig3}(d).
Figures~\ref{fig3}(e) and \ref{fig3}(f) show the temperature dependence of critical current normalized by critical current at 0~K limit, $I_{c,sf}(T)/I_{c,sf}(0)$, down to 50~mK. Since $I_{c,sf}(T)$ is essentially temperature-independent below $\sim$1.5~K, we take $I_{c,sf}(50~\rm{mK})$ as $I_{c,sf}(0)$. The experimental data are shown as solid symbols. When the temperature is reduced, $I_{c,sf}(T)/I_{c,sf}(0)$ first increases significantly and then rapidly saturates at the zero temperature limit, which we will show to be consistent with a nodeless order parameter~\cite{Talantsev2015, Talantsev2017, Liu2022}. 

\begin{figure*}[!t]\centering
      \resizebox{18.5cm}{!}{
 \includegraphics{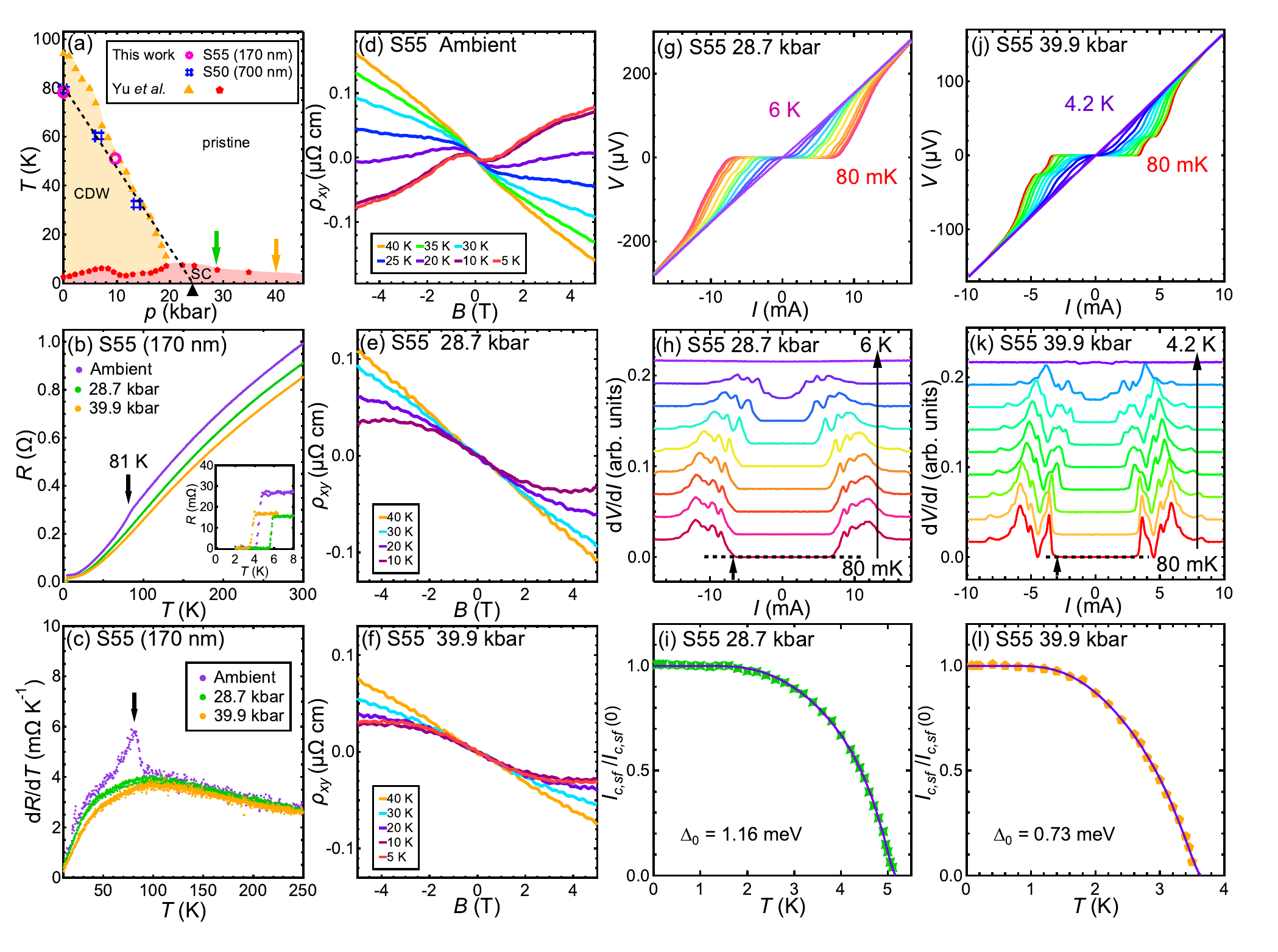}}       \caption{\label{fig4} 
 \textbf{(a)} Temperature-pressure phase diagram of~\CsVSb. The solid symbols are adapted from Ref.~\cite{Yu2021}, where the CDW phase is totally suppressed at around 20~kbar. The open symbols represent the $T_{\rm CDW}$ in thin flakes measured by us, where the CDW phase is suppressed at around 24~kbar based on linear extrapolation, as indicated by the up-triangle. The green and yellow arrows at 28.7 kbar and 39.9 kbar, respectively, indicate the pressure values chosen for investigating superconductivity free of TRSB. 
 \textbf{(b)} Temperature dependence of resistance of S55 (170 nm) at ambient pressure, 28.7 kbar and 39.9 kbar. The inset shows the superconducting transition.
 \textbf{(c)} Temperature dependence of ${\rm d}R/{\rm d}T$ of S55 at ambient pressure, 28.7 kbar and 39.9 kbar. The arrow indicates $T_{\rm CDW}$.
 \textbf{(d)} Hall resistivity $\rho_{xy}$ of S55 at different temperatures at \textbf{(d)} ambient pressure, \textbf{(e)} 28.7 kbar and \textbf{(f)} 39.9 kbar. $V$-$I$ characteristics of S55 at different temperatures at \textbf{(g)} 28.7 kbar and \textbf{(j)} 39.9 kbar. The calculated first derivative of $V(I)$, d$V$/d$I$, of S55 at \textbf{(h)} 28.7 kbar and \textbf{(k)} 39.9 kbar. The short black arrows indicate the onset of deviation from  d$V$/d$I=0$, which we define as the critical current $I_c$. The long arrows show the direction of increasing temperature. Temperature dependence of $I_{c,sf}(T)/I_{c,sf}(0)$ of S55 at \textbf{(i)} 28.7 kbar and \textbf{(l)} 39.9 kbar. The solid curves are 
 \leftline{the single $s$-wave gap fits.}
 }    
\end{figure*}

When the half-thickness ($b$) of the flake is smaller than the penetration depth ($\lambda$), the self-field critical current density ($J_{c,sf}$), i.e. the transport critical current density at the zero magnetic field, is recently established to relate to the penetration depth $\lambda$ as follows~\cite{Talantsev2015}:
\begin{equation}
J_{c,sf}=\frac{\phi_0}{4\pi \mu_0\lambda^3}\left(\ln\frac{\lambda}{\xi}+0.5\right)
\label{Jc_eqn}
\end{equation}
where $\phi_0$ is the flux quantum and $\xi$ is the coherence length. Since the superfluid density $\rho_s \propto \lambda^{-2}$, the temperature dependence of $\rho_s$ can be determined by a careful measurement of $J_{c,sf}(T)$, allowing a discussion of the superconducting gap. In particular, for $s$-wave symmetry with a single gap,
\begin{equation}
\frac{\rho_s(T)}{\rho_s(0)}=\frac{\lambda(T)^{-2}}{\lambda(0)^{-2}}=1-2\sqrt{\frac{\pi\Delta(0)}{k_BT}}e^{-\Delta(0)/k_BT}.
\label{gap_eqn}
\end{equation}

From the muon spin rotation measurement, the penetration depth of \CsVSb\ is around 450~nm in the low temperature limit~\cite{Gupta2022}. Thus, Eqn.~(\ref{Jc_eqn}) is applicable to both M1 and M4 as both flakes satisfy the condition $b\ll\lambda$. 
In fact, $I_{c,sf}(T)/I_{c,sf}(0)=J_{c,sf}(T)/J_{c,sf}(0)$, because the sample geometry is temperature independent and the terms enclosed in the brackets in Eqn.~(\ref{Jc_eqn}) can be regarded as constants because of the logarithm.  Thus, we can use the combination of Eqns.~(\ref{Jc_eqn}) and (\ref{gap_eqn}) to analyze our data. As shown in Figs.~\ref{fig3}(e) and \ref{fig3}(f), the experimental data can be accurately described by assuming an $s$-wave gap (solid curves). Besides, the extracted superconducting gap values are 0.95 meV (2.70 $k_BT_c$) and 1.08 meV (2.84 $k_BT_c$) for M1 and M4, respectively, which are larger than the BCS weak coupling limit (1.76 $k_BT_c$), indicating strong-coupling superconductivity. The strong coupling nature of the superconductivity revealed here is consistent with previous studies (see Supporting Information S3 for a comparison with the literature values). Figure~\ref{fig3}(g) displays $I_{c,sf}(T)/I_{c,sf}(0)$ versus the reduced temperature $t=T/T_c$ for all thin flakes we study and all the datasets collapse into a single curve, indicating both the consistency of our results and the preservation of the same order parameter down to the lowest temperature.

To investigate the possible influence of TRSB -- introduced via the CDW order -- on the superconducting gap, we take advantage of the known temperature-pressure phase diagram constructed for \CsVSb. For the bulk system, the CDW order can be completely suppressed by a pressure of $\sim$20~kbar while in the thin flake, the CDW order disappears at $\sim$24~kbar (Fig.~\ref{fig4}(a)). Thus, we performed two experiments at 28.7~kbar and 39.9~kbar to examine the superconducting state without the complication due to TRSB. The absence of the CDW order at 28.7~kbar and 39.9~kbar is evidenced in $R(T)$ and d$R$/d$T$, where the signature of the CDW is absent at high pressure, in sharp contrast to the ambient pressure data for the same flake (S55) (see Figs.~\ref{fig4}(b) and \ref{fig4}(c)). {In addition, the residual resistance values at 28.7~kbar and 39.9~kbar are comparable, but noticeably lower than that at ambient pressure (inset of Fig.~\ref{fig4}(b)). This is because at 28.7~kbar and 39.9~kbar, the elimination of the CDW state implies the absence of the CDW-induced Fermi surface gapping, giving rise to a smaller resistance.

Further evidence for the removal of the CDW state at high pressure is provided by the absence of the anomalous Hall effect. Our $\rho_{xy}$ data at ambient and high pressure strongly resembles the data reported by Yu \etal~\cite{Yu2021b}: the characteristic `$S$ shape' line in the low-field region at ambient pressure attributable to AHE is also detected. At 28.7~kbar and 39.9~kbar, the curvature of $\rho_{xy}(B)$ varies more slowly, which can instead be described with a two-band model. Following Yu \etal, the disappearance of the AHE is tied to the removal of the CDW state.

We proceed to the measurement of the critical current under pressure. Figure~\ref{fig4}(g) shows the $V$-$I$ curves at selected temperatures at 28.7~kbar. Following the same procedure, we extract $I_{c,sf}(T)/I_{c,sf}(0)$ from d$V$/d$I$ (see Fig.~\ref{fig4}(h)), allowing a glimpse of the superconducting gap in \CsVSb\ without the accompanying CDW order. At 28.7~kbar,  $I_{c,sf}(T)/I_{c,sf}(0)$ can again be described by Eqns.~(\ref{Jc_eqn}) and (\ref{gap_eqn}), indicating an $s$-wave gap (Fig.~\ref{fig4}(i)). At 39.9~kbar, which is $\sim$1.6 to 2 times higher than the critical pressure at which $T_{\rm CDW}$ extrapolates to zero, we again obtain results consistent with an $s$-wave gap (see Figs.~\ref{fig4}(j)-(l)). Note, however, the complicated structure in d$V$/d$I$ at 39.9~kbar, and the existence of a sharp dip in d$V$/d$I$ beyond $I_c$. This can be caused by inhomogeneity, or the existence of another gap. Indeed, the analysis of the dip results in a gap-like temperature dependence with a $s$-wave symmetry (see Supporting Information S2). We tentatively remark that a second superconducting gap is possible in \CsVSb. Nevertheless, these results unambiguously show that the CDW state does not modify the symmetry of the superconducting gap. This is in sharp contrast to the sister compound RbV$_3$Sb$_5$, in which $\mu$SR detects a transition from a nodal to a nodeless gap when the CDW state is suppressed by pressure~\cite{Guguchia2022}.

One aspect that is affected by the CDW order in \CsVSb\ is the coupling strength, as benchmarked by the dimensionless ratio 2$\Delta/k_BT_c$. The superconducting gap is 1.16~meV and 0.73~meV at 28.7~kbar and 39.9~kbar, respectively. These gap values gives 2$\Delta/k_BT_c$ of 5.20 and 4.66, both smaller than the smallest value 5.30 at ambient pressure (Fig.~\ref{fig3}(g)). Nevertheless, 2$\Delta/k_BT_c$ are all higher than the BCS weak-coupling limit over the pressure range we investigated. Thus, superconductivity in \CsVSb\ is strong-coupling, but the coupling strength appears to be sensitive to pressure.

The observed conventional $s$-wave superconductivity is consistent with other CDW systems~\cite{Goh2015,Yu2015,Gruner2017}, and the evolution of the coupling strength is reminiscent of (Sr,Ca)$_3$Rh$_4$Sn$_{13}$, in which the coupling strength is progressively enhanced towards the structural/CDW quantum critical point~\cite{Yu2015}. The enhancement of the coupling strength in (Sr,Ca)$_3$Rh$_4$Sn$_{13}$ has been traced to the softening of a phonon mode associated with the second-order structural transition~\cite{Cheung2018}. Recently, phonon softening has also been reported in Lu(Pt$_{1-x}$Pd$_x$)$_2$In, another superconducting system with a CDW transition tunable by $x$~\cite{Gruner2022}. In \CsVSb, the suppression of the CDW state may result in a quantum critical point. On approaching the quantum critical point from either side, part of the phonon spectrum would gradually be softened, leading to an enhanced 2$\Delta/k_BT_c$. This scenario appears to explain our observation, as both 28.7~kbar and 39.9~kbar are beyond the CDW region and 2$\Delta/k_BT_c$ shows a clear decreasing trend as the system moves away from the CDW phase. However, detailed studies are still needed to examine how $2\Delta/k_BT_c$ varies within the CDW phase. Recent first-principles density-functional theory calculations indeed reveal the softening of phonon modes around the $L$ point of the Brillouin zone upon approaching the CDW phase from high pressure~\cite{Wang2022}, lending support to our experimental results.   
~\\

\begin{center}{\bf METHODS}\end{center}
\textit{Crystal growths.}
Single crystals of \CsVSb~were synthesized from Cs (ingot, 99.95 $\%$), V (powder, 99.9 $\%$) and Sb (shot, 99.9999 $\%$) using self-flux method similar to Refs.~\cite{Ortiz2019,Ortiz2020}. The cooling rate of the final segment of the growth profile was adjusted to prepare samples with varying RRR values. For example, the best sample comes from the growth where the final segment was cooling from 900$^\circ$C to room temperature at 0.5$^\circ$C/hr, while for the sample with RRR$<$30, the corresponding cooling rate was 2$^\circ$C/hr. \\
\\
\textit{High pressure.}
We adopt the concept of ``device integrated diamond anvil cell" developed by us for high-pressure studies~\cite{Xie2021,Ku2022}. Our anvils were patterned with microelectrodes by photo-lithography and physical vapour deposition coating. Thin flakes of \CsVSb\ were exfoliated from single crystals and then transferred onto the patterned electrodes. A thin layer of h-BN was added onto the thin flakes for encapsulation. The thickness of the thin flakes was determined by a dual-beam focused ion beam system (Scios~2 DualBeam by Thermo Scientific) prior to pressurization. High-purity glycerine 99.5\% was used as the pressure transmitting medium. The pressure achieved was determined by ruby fluorescence at room temperature~\cite{Mao1986}.\\
\\
\textit{Measurements.}
Electrical resistivity was measured by a standard four-terminal configuration in Physical Property Measurement System by Quantum Design and a dilution refridgerator by Bluefors. 
Dupont 6838 silver paste was used for making the electrical contacts on bulk crystals, while a set of patterned electrodes was used to form a tight contact with thin flakes exfoliated from the bulk crystals~\cite{Xie2021,Ku2022}. $V$-$I$ curves were measured by a Keithley 2182A nanovoltmeter together with a Keithley 6221 current source in the pulsed delta mode. The duration of the pulsed current was 11~ms, and the pulse repetition time was 1~s. 
\\

\begin{center}{\bf ASSOCIATED CONTENT}\end{center}
\noindent{\bf Supporting Information}\\
The Supporting Information is available free of charge via the internet at http://pubs.acs.org.\\
Figures S1 and S2 display additional $\rho(T)$ data and Fig. S3 shows the RRR dependence of $T_{\rm CDW}$. Figure S4 is the analysis of the dip feature in d$V$/d$I$ at 39.9 kbar. Table S1 displays the comparison of the superconducting gap magnitude with previous studies.

\begin{center}{\bf AUTHOR INFORMATION}\end{center}
{\bf Corresponding Authors}\\
$^*$Kwing To Lai. E-mail: ktlai@phy.cuhk.edu.hk\\
$^{\dagger}$Swee K. Goh. E-mail: skgoh@cuhk.edu.hk\\

\noindent{\bf Author Contributions}\\
\noindent S.K.G. proposed and supervised the project. W.Z. coordinated the experimental efforts. W.Z., X.L., L.W. and W.W. conducted measurements, C.W.T., S.T.L. and K.T.L. prepared single crystals of \CsVSb. J.X. provided assistance with pressure cells. X.Z., Y.Z. and S.W. provided h-BN. J.L.T. provided theoretical support. W.Z., Z.W. and S.K.G. performed data analysis. W.Z., K.T.L. and S.K.G. wrote the manuscript with input from all authors. 
\\

\noindent{\bf Notes}\\
The authors declare no competing financial interest.\\
\\
\leftline{\bf Data availability}
\noindent All the data that support the findings of this paper are available from the corresponding authors upon reasonable request.
\\

\begin{center}{\bf ACKNOWLEDGMENTS}\end{center}
\noindent The authors acknowledge insightful comments from Yajian Hu and Chongze Wang.
This work was supported by Research Grants Council of Hong Kong (CUHK 14301020, CUHK 14300722, A-CUHK402/19), CUHK Direct Grant (4053461, 4053408, 4053528, 4053463), the National Natural Science Foundation of China (12104384, 12174175) and the Shenzhen Basic Research Fund (JCYJ20190809173213150). \\


\providecommand{\noopsort}[1]{}\providecommand{\singleletter}[1]{#1}%

\end{document}